\documentclass{iopart}
\usepackage{iopams}
\usepackage{graphicx}
\usepackage{harvard}
\usepackage{epstopdf}
\usepackage{verbatim}
\usepackage{setstack}
\usepackage{wasysym}
\usepackage[thickspace,amssymb]{SIunits}

\begin{document}
\title[]{Charged particle's flux measurement from PMMA irradiated by 80 MeV/u carbon ion beam}

\author{C.~Agodi$^{f}$, G.~Battistoni$^{g}$, F.~Bellini$^{a,b}$, G.A.P.~Cirrone$^{f}$, F.~Collamati$^{a,b}$, G.~Cuttone$^{f}$, E.~De Lucia$^{c}$, M.~De Napoli$^{f}$, A.~Di Domenico$^{a,b}$, R.~Faccini$^{a,b}$, F.~Ferroni$^{a,b}$, S.~Fiore$^{a}$, P.~Gauzzi$^{a,b}$, E.~Iarocci$^{c,d}$, M.~Marafini$^{a,e}$, I.~Mattei$^{a,b}$, S.~Muraro$^{g}$, A.~Paoloni$^{c}$, V.~Patera$^{c,d}$, L.~Piersanti$^{c,d}$, F.~Romano$^{f}$, A.~Sarti$^{c,d}$, A.~Sciubba$^{c,d}$, E.~Vitale$^{g}$, C.~Voena$^{a,b}$}
\address{$^a$ Dipartimento di Fisica, Sapienza Universit\`a di Roma, Roma, Italy}
\address{$^b$ INFN Sezione di Roma, Roma, Italy}
\address{$^g$ INFN Sezione di Milano, Milano, Italy}
\address{$^c$ Laboratori Nazionali di Frascati dell'INFN, Frascati, Italy}
\address{$^d$ Dipartimento di Scienze di Base e Applicate per Ingegneria, Sapienza Universit\`a di Roma,  Roma, Italy}
\address{$^e$ Museo Storico della Fisica e Centro Studi e Ricerche ``E.~Fermi'', Roma, Italy}
\address{$^f$ Laboratori Nazionali del Sud dell'INFN, Catania, Italy}

\begin{abstract}
Hadrontherapy is an emerging technique in cancer therapy that uses beams of charged particles.
To meet the improved capability of hadrontherapy in matching the dose release with the cancer position, new dose monitoring techniques need to be developed and introduced into clinical use. The measurement of the fluxes of the secondary particles produced by the hadron beam is of fundamental importance in the design of any dose monitoring device and is eagerly needed to tune Monte Carlo simulations.
We report the measurements done with charged secondary particles produced from the interaction of a 80 MeV/u fully stripped carbon ion beam at the INFN Laboratori Nazionali del Sud, Catania, with a  Poly-methyl methacrylate target. Charged secondary particles, produced at 90$\degree$ with respect to the beam axis, have been tracked with a drift chamber, while their energy and time of flight has been measured by means of a LYSO scintillator. Secondary protons have been identified exploiting the energy and time of flight information, and their emission region has been reconstructed backtracking from the drift chamber to the target. Moreover a position scan of the target indicates that the reconstructed emission region follows the movement of the expected Bragg peak position. Exploting the reconstruction of the emission region, an accuracy on the Bragg peak determination in the submillimeter range has been obtained. The measured differential production rate for protons produced with $E^{\rm Prod}_{\rm kin} >$  83 MeV and emitted at 90$\degree$ with respect to the beam line is: $dN_{\rm P}/(dN_{\rm C}d\Omega)(E^{\rm Prod}_{\rm kin} > 83 {\rm ~MeV},  \theta=90\degree)= (2.69\pm 0.08_{\rm stat} \pm 0.12_{\rm sys})\times 10^{-4} sr^{-1}$.
\end{abstract}
\vspace{2pc}
\noindent
{\it Keywords\/}: drift chamber; LYSO; hadrontherapy; carbon ion beam, dose monitoring 
\noindent
\submitto{\PMB}
\maketitle
\section{Introduction}
Protons and carbon ion beams are presently used to treat many different solid cancers \cite{Jakel2008,Durante2010} and several new centers based on hadron accelerators are operational or under construction \cite{Amaldi2005,Schardt2010}.
The main advantage of this technique, in comparison to the standard radiotherapy with X-ray beams, is the better localization of the irradiation dose in the tumor affected region sparing healthy tissues and possible surrounding organs at risk. This feature can be achieved because the heavy charged particles loose most of the energy at end of their range, the Bragg peak (hereafter BP), in comparison to the exponentially decreasing energy release of the X-ray beam. Up to now most of the patients have been treated at centers with proton beams, but routinary use of carbon beams has now started. There are also proposals for future use of $^{4}He$, $^{7}Li$ or $^{16}O$ beams \cite{Brahme1986}.\\
New dose monitoring techniques need to be developed and introduced into clinical use, to meet the improved capability of hadrontherapy to match the dose release with the cancer position. The R\&D effort should be then focused to develop novel imaging methods to monitor, preferably in real time, the 3-dimensional distribution of the radiation dose effectively delivered during hadrontherapy. \\
 This holds true especially for treatments using carbon ion beams since the dose profile is very sensitive to anatomical changes and minor patients' positioning uncertainties.  \\
Conventional methods for the assessment of patients' positioning used in all X-ray based radiation therapy, where a non-negligible fraction of the treatment beam is transmitted through the patient, cannot be used to pursue this task due to the different physics underlying.
All the proposed methods exploit the information provided by the secondary particles produced by the hadron beam along its path to the tumour, inside the patient's body. In particular it has been already shown that the peak of the dose released by the hadron beam can be correlated with the emission pattern of the flux of secondary particles created by the beam interaction, namely: i) prompt photons within the 1-10 MeV energy range~\cite{Min2006,Testa2008,Testa2009} and ii) pairs of back-to-back photons produced by the annihilation of positrons coming from $\beta^+$ emitters, mainly $^{11}C$ and $^{15}O$~\cite{Pawelke1997,Parodi2002,Enghardt2004,Fiedler2008,Vecchio2009,Attanasi2009}.\\
In this paper we suggest the possibility to correlate the position of the BP in the patient with the emission region of charged secondary particles, mainly protons with kinetic energy E$_{\rm kin}$<150 MeV. 
We report the study of the charged secondary particles produced from the irradiation of a Poly-methyl methacrylate (PMMA) with the 80 MeV/u fully stripped carbon ion beam of the INFN Laboratori Nazionali del Sud (LNS). Section~\ref{setup} is devoted to the description of the experimental setup;
the event selection and the spectra of the charged secondary particles are presented in Section~\ref{datasel}. The analysis of the production region of charged secondary particles is described in Section~\ref{region}, and the measurement of their differential production rate is reported in Section~\ref{flux}.
\section{Experimental setup}
\label{setup}

\begin{figure}[!ht]
\begin{center}
\centerline{
\includegraphics [width = 1 \textwidth] {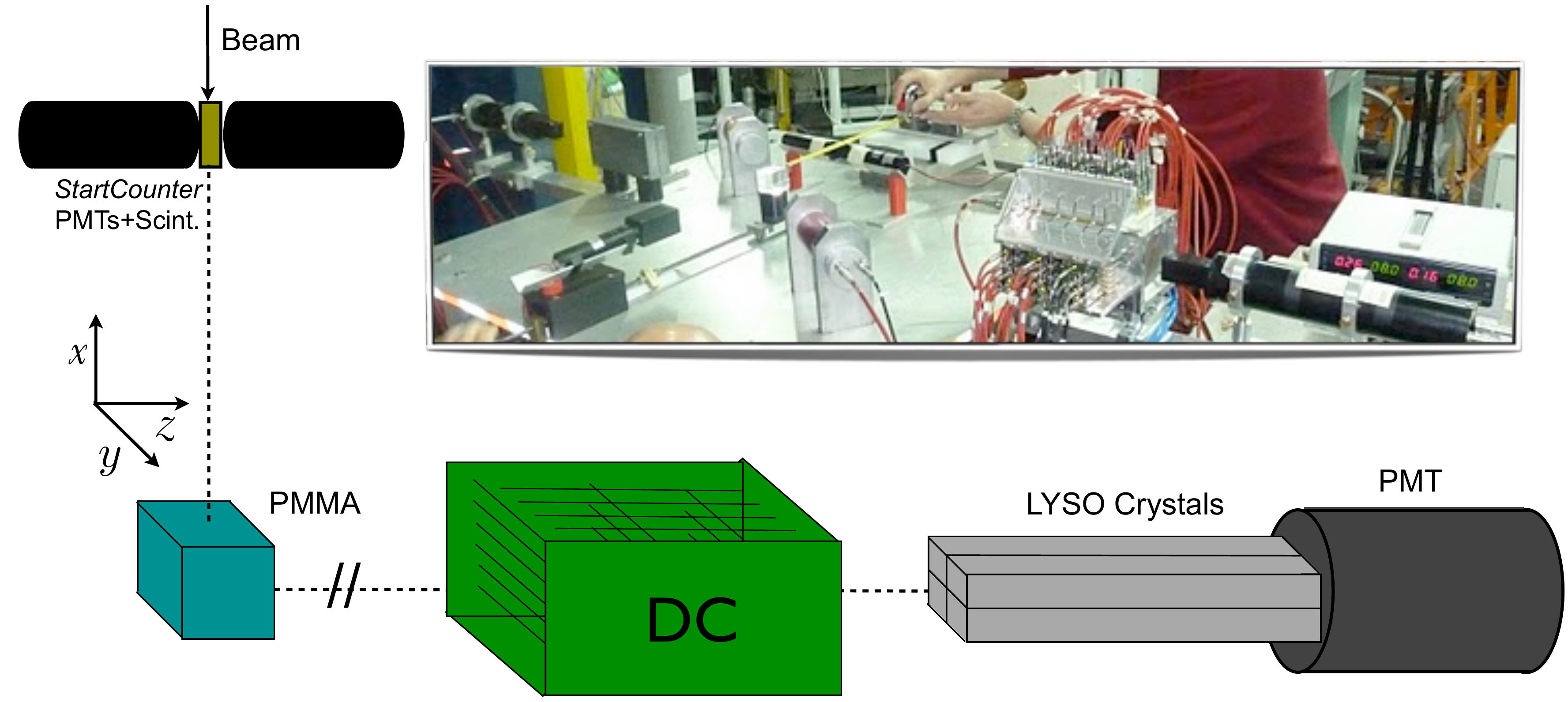}}
\caption{\small{Experimental setup: picture and schematic view. The acquisition is triggered by the coincidence of the Start Counter and the LYSO detector.}}
\label{fig:Schema}
\end{center}
\end{figure}
Figure~\ref{fig:Schema} shows the experimental setup.
A 4x4x4 cm$^3$ PMMA target is placed on a 80 MeV/u, fully stripped $^{12}C$ ion beam. The beam rate, ranging from hundreds of kHz to $\sim 2$ MHz, is monitored with a 1.1 mm thick scintillator placed at 17 cm from the PMMA on the beam line and read-out by two photomultiplier tubes (PMTs) Hamamatsu 10583 put in coincidence (Start Counter). \\
An array of 4 LYSO crystals, each measuring 1.5x1.5x12 cm$^3$, is placed at 90$\degree$ with respect to the beam line, at 74 cm from the PMMA center. The scintillation light of the crystals is detected with a PMT EMI 9814B triggered in coincidence with the Start Counter.

\begin{figure}[!ht]
\begin{center}
\includegraphics [width = 1\textwidth] {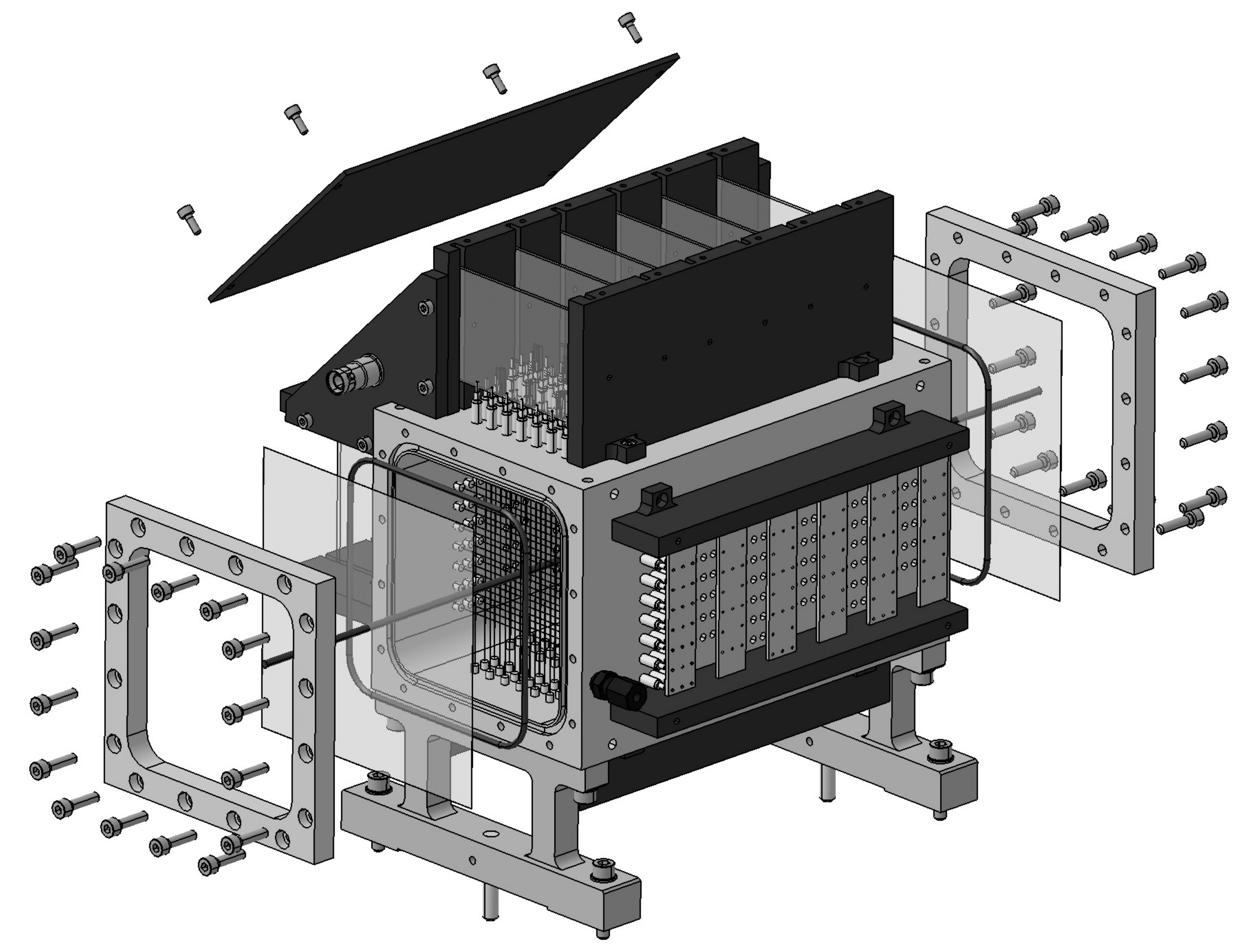}
\caption{\small{Mechanical drawing of the drift chamber.}}
\label{fig:Camera}
\end{center}
\end{figure}

\noindent
A 21 cm long drift chamber~\cite{Abou-Haidar2012} is placed at 51 cm from the PMMA center, along the flight line connecting the PMMA to the LYSO crystals. 
We have chosen the configuration at 90$\degree$ with respect to the beam line to maximize the sensitivity to the Bragg peak position along the beam.
In the following the coordinate system is defined (Figure~\ref{fig:Schema}) with the x-axis along the beam line toward the Start Counter, the z-axis along the line connecting the centers of PMMA, drift chamber and LYSO crystals detector and the y-axis oriented according to the right-hand rule, and origin at the drift chamber center. The PMMA center in the vertical axis was at y = -4 mm.
The drift chamber provides a 2-dimensional reconstruction of the space point by alternated horizontal (x-z plane V-view) and vertical (y-z plane U-view) layers of wires.  Each layer is composed of three 16x10 mm$^2$ rectangular cells for a total of 36 sense wires (Figure~\ref{fig:Camera}). The twelve layers, six on each view, provide tracking redundancy and ensure high tracking efficiency and excellent spatial resolution. In order to minimize tracking ambiguities, the consecutive layers of each view are staggered by half a cell. 
Custom front-end electronics boards, designed and realized at the INFN Laboratori Nazionali di Frascati (LNF) electronics workshop, are embedded in the detector and provide single wire signal amplification by a factor of 10. The drift chamber has been operated with 1.8 kV sense wire voltage, Ar/CO$_2$ (80/20) gas mixture and 30 mV discriminating threshold for the signals, achieving  $\leq 200~\mu$m single cell spatial resolution and $\simeq 96\%$ single cell efficiency~\cite{Abou-Haidar2012}.

\noindent
The signals from the Start Counter and the LYSO crystals are split and fed into a 12-bit QDC (Caen V792N) and a 19-bit TDC (Caen V1190B), after discrimination, to provide the measurements of both the particles' energy and arrival time. The signals from the 36 cells of the drift chamber are fed, after discrimination, into the same TDC providing the drift time measurements.  The front-end electronics has been read-out by a VME system using a MOTOROLA 5100 CPU board.

\noindent The energy and time calibration of the LYSO crystals and the determination of the drift chamber space-time relations have been described in \cite{Agodi2012a,Agodi2012} and \cite{Abou-Haidar2012} respectively. A custom tracking algorithm, based on a least squares iterative fitting method, has been also developed to reconstruct the direction of the charged secondary particles. A first track reconstruction is performed using very clean topologies asking for at least three layers with a single fired cell (hit), on both the V- and U-views. Then a hit addition algorithm improves the tracking performance by using the information from all the other layers.

\section{Data selection}
\label{datasel}
The trigger signal is provided by the coincidence of the Start Counter and LYSO crystals signals, within 80 ns. 
\begin{figure}[!htb]
\includegraphics [width = \textwidth] {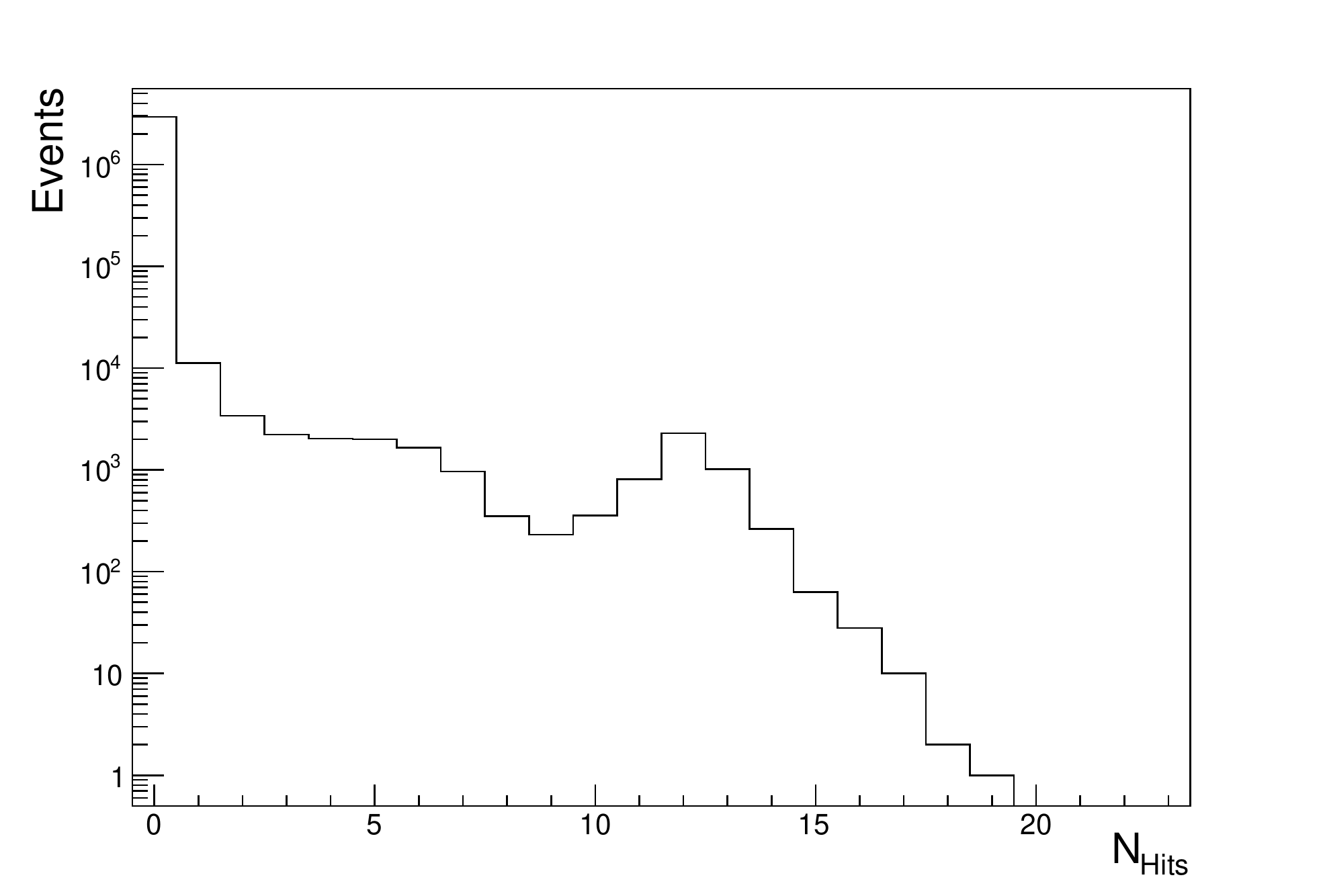}
\caption{\small{Distribution of the number of fired cells in the drift chamber $N_{\rm Hits}$ for events with detected energy in the LYSO crystals $E_{ \rm LYSO} >$ 1 MeV.}}
\label{fig:Nhits}
\end{figure}
Figure~\ref{fig:Nhits} shows the distribution of the number of hits  in the drift chamber ( $N_{\rm Hits}$ ) obtained for events with detected energy in the LYSO crystals $E_{\rm LYSO} > $  1 MeV. Events with $N_{\rm Hits} >$ 9 are selected for the analysis of the charged secondary particles.

\begin{figure}[!h]
\begin{center}
\includegraphics [width = 1\textwidth] {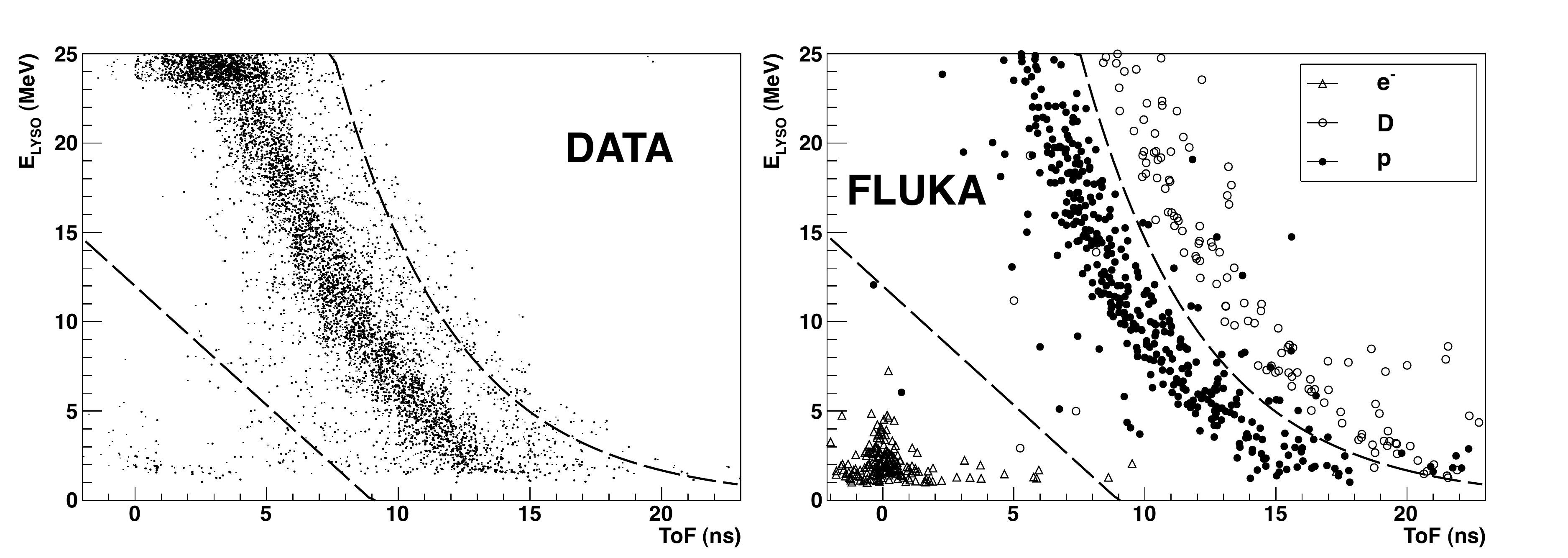}
\caption{\small{ Distribution of the detected energy in the LYSO crystals as a function of the Time of Flight: Data (Left) and FLUKA Simulation (Right).
}}
\label{fig:Etof}
\end{center}
\end{figure}

\noindent
In order to evaluate the setup acceptance and efficiency, and to optimize the particle identification analysis a detailed simulation has been developed using the FLUKA software release 2011.2~\cite{Fasso'2003,Ferrari2005}. The detailed geometry description with the setup materials (air included) together with the trigger logic, the time resolution of the scintillator as well as the experimental space resolution of the drift chamber have been considered. The quenching effect in the scintillator has also been introduced in the Monte Carlo according to \cite{Koba2011}.
The interaction of a sample of $10^9$ carbon ions with 80 MeV/u, equivalent to $10^3$ s of data taking at the typical 1 MHz rate of beam, has been simulated. To identify the charged particles reconstructed in the drift chamber, we exploit the distribution of the detected energy in the LYSO detector $E_{\rm LYSO}$  as a function of Time of Flight (ToF), Figure~\ref{fig:Etof}. 
In the data sample (left panel) a fast low-energy component due to electrons
is clearly visible for ToF values around zero, in the area delimited by the first dashed line. These electrons are produced by Compton scattering of the de-excitation photon induced by beam interactions in the PMMA material. 
The central most populated band, delimited by the two dashed lines, is made by protons with detected energy within a very wide range, originating also the clearly visible saturation of the LYSO crystals QDC for $E_{\rm LYSO} >$ 24 MeV. The FLUKA simulation (right panel) shows similar populations in the (ToF , $E_{\rm LYSO}$) plane with an additional component of deuterons, above the second dashed line, which is not present in data.

\begin{figure}[!htb]
\begin{center}
\includegraphics [width = \textwidth] {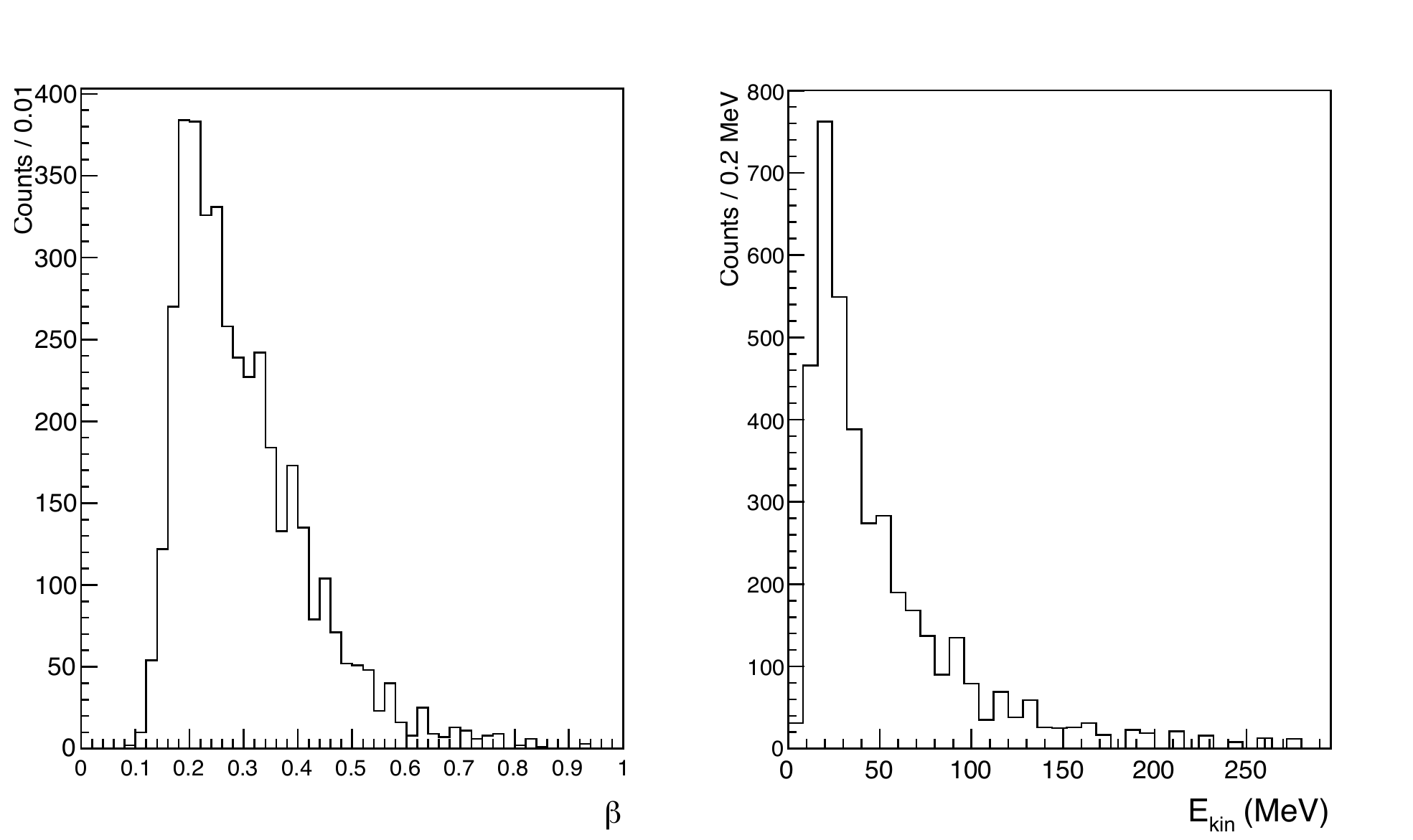}
\caption{\small{Distribution of $\beta = \frac{v}{c}$ (left) and kinetic energy (right) of charged secondary particles identified as protons.}}
\label{fig:beta}
\end{center}
\end{figure}

\noindent
We have then identified as proton a charged secondary particle with ToF and $E_{\rm LYSO}$ values inside the area delimited by the two dashed lines in Figure~\ref{fig:Etof}. 
The systematic uncertainty on the proton/deuteron identification has been estimated using the data events in the deuterons area of the (ToF , $E_{\rm LYSO}$) plane. 
Figure~\ref{fig:beta} shows the distributions of $\beta = \frac{v}{c}$ and the corresponding detected kinetic energy $E_{\rm kin}$ for the identified protons, obtained using the ToF measurement together with the distance between LYSO crystals and PMMA.
This detected kinetic energy can be related to the proton kinetic energy at emission time, $E^{\rm Prod}_{\rm kin}$, considering the energy loss in the PMMA and the quenching effect of the scintillating light for low energy protons. The minimum required energy to detect a proton in the LYSO crystals is $E^{\rm Prod}_{\rm kin} = 7.0\pm 0.5$ MeV, evaluated using the FLUKA simulation, and a proton with an average detected kinetic energy $E_{\rm kin}$ = 60 MeV has been emitted with $E_{\rm kin}^{\rm Prod} 83\pm 5$ MeV. The uncertainty is mainly due to the finite size of both the beam spot $\mathcal{O}$(1 cm) and profile.

\noindent
In order to use the secondary protons for monitoring purposes, the crossing of some centimeters of patient's tissue has to be considered and therefore the range  $E_{\rm kin} > $ 60 MeV of the detected kinetic energy distribution is the most interesting for the above-mentioned application.
In the following the proton kinetic energy detected in the LYSO crystals will be referred to as the kinetic energy.
\section{Production region of charged secondary particles}
\label{region}
Tracks reconstructed in the drift chamber are backward extrapolated to the PMMA position, to find the production region of charged secondary particles along the path of the carbon ion beam. The PMMA is mounted on a single axis movement stage allowing position scans along the x-axis to be performed with a 0.2 mm accuracy (Figure~\ref{fig:Schema}). In the configuration with the centers of PMMA, drift chamber and LYSO crystals aligned along the z-axis, the PMMA position in the stage reference frame is taken as 0 and will be referred to as the reference configuration. 

\begin{figure}[!htb]
\begin{center}
\includegraphics [width = 0.8\textwidth] {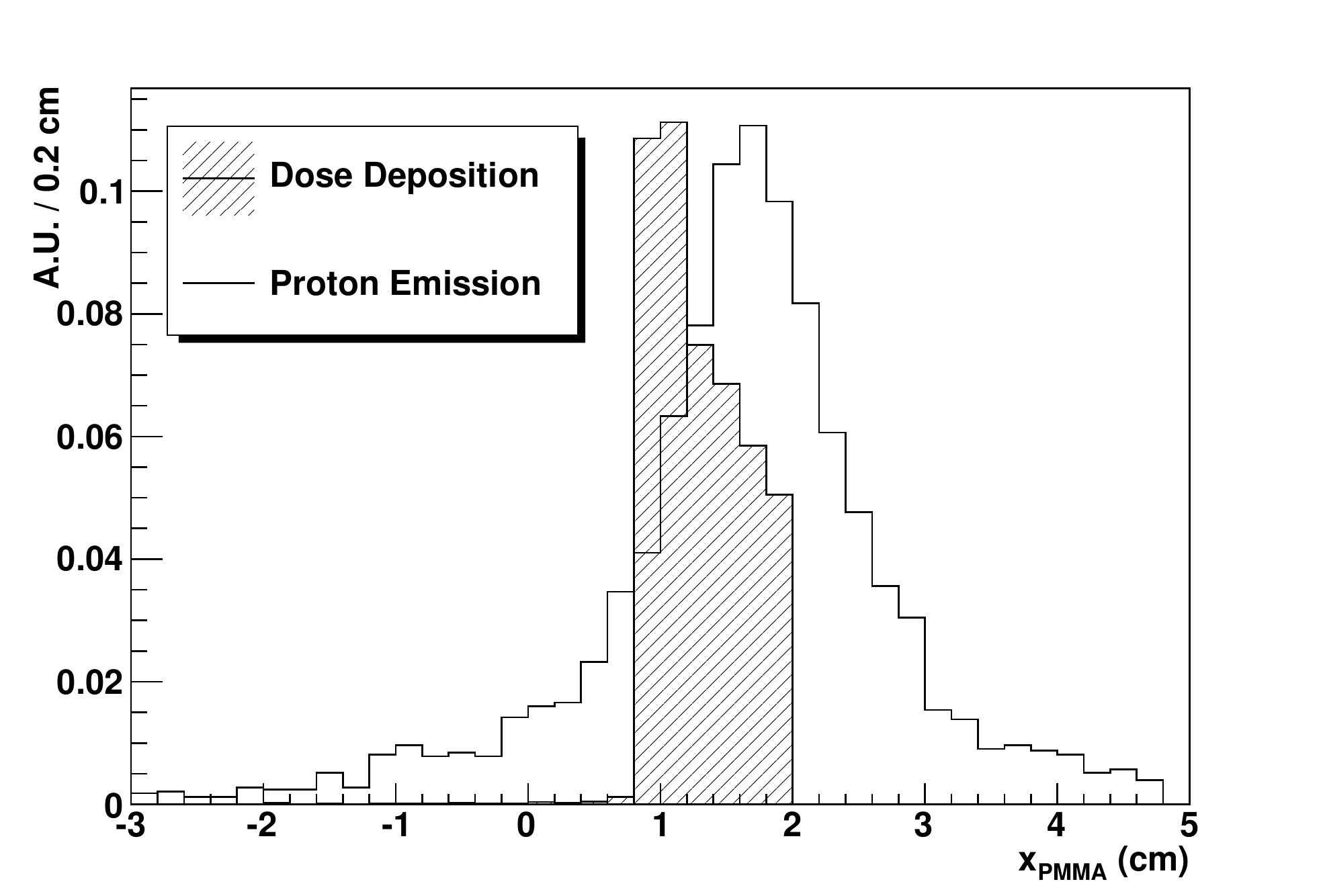}
\caption{\small{Expected dose deposition in the PMMA  evaluated with FLUKA (hatched) compared to the distribution of $x_{\rm PMMA}$ (solid), the emission point of charged secondary particles along the x-axis. The beam entrance and exit faces of the PMMA are at $x_{\rm PMMA}$ = 2 cm and $x_{\rm PMMA}$ = -2 cm, respectively.}}
\label{fig:peak}
\end{center}
\end{figure}

\begin{figure}[!ht]
\begin{center}
\includegraphics [width = 1 \textwidth] {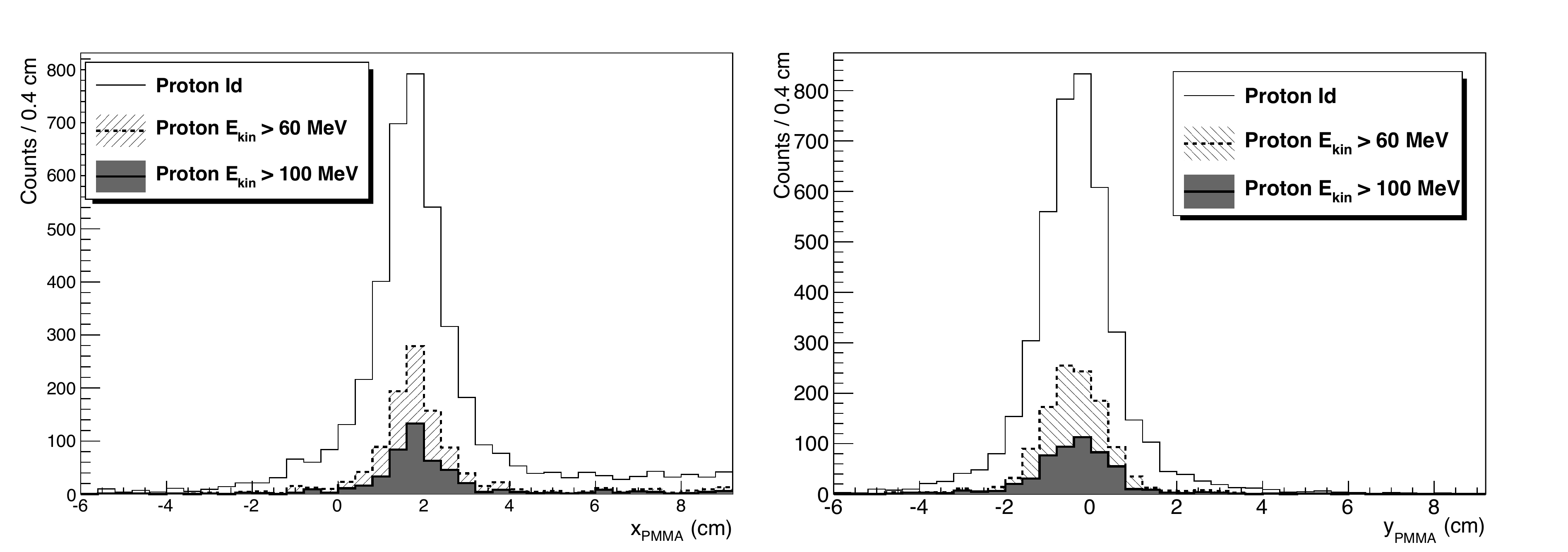}
\caption{\small{Distribution of $x_{\rm PMMA}$ (Left) and $y_{\rm PMMA}$ (Right) obtained for all charged particles identified as protons (black solid line), for protons with $E_{\rm kin} > $ 60 MeV (dashed line) and with $E_{\rm kin} > $ 100 MeV (grey).  The beam entrance and exit faces of the PMMA are at $x_{\rm PMMA}$ = 2 cm and $x_{\rm PMMA}$ = -2 cm, and $y_{\rm PMMA}$ = 1.6 cm and $y_{\rm PMMA}$ = -2.4 cm.}}
\label{fig:reso_ekin}
\end{center}
\end{figure}

\noindent From each track reconstructed in the drift chamber and backward extrapolated to the beam axis we can measure the x and y coordinates of the estimated emission point of the charged secondary particle, named $x_{\rm PMMA}$ and $y_{\rm PMMA}$. The expected position of the Bragg peak obtained with the FLUKA simulation \cite{Fasso'2003} is located at $(11.0 \pm 0.5)$ mm from the beam entrance face of the PMMA. With the setup in the reference configuration, the expected position of the Bragg peak in our coordinate system is $x_{\rm Bragg}|^{\rm Ref} = (9.0 \pm 0.5)$ mm. 
Figure~\ref{fig:peak} shows the distribution of the reconstructed $x_{\rm PMMA}$, compared to the expected distribution of the dose deposition in the PMMA, both obtained with the setup in the reference configuration. 
The mean of the gaussian fit to the distribution is $ \bar{x}_{\rm PMMA} = 17.1\pm0.2$ mm, and consequently the separation between the BP and the peak from secondary proton emission is $\Delta_{\rm ProtonBragg} = 8.1 \pm 0.5$ mm.
 Figure~\ref{fig:reso_ekin} shows the distribution of the reconstructed $x_{\rm PMMA}$ and $y_{\rm PMMA}$ for all identified protons (solid line), for protons with $E_{\rm kin} > $ 60 MeV (hatched) and for protons with $E_{\rm kin} >$ 100 MeV (grey). The beam entrance and exit faces of the PMMA are at $x_{\rm PMMA}$ = 2 cm and $x_{\rm PMMA}$ = -2 cm, and $y_{\rm PMMA}$ = 1.6 cm and $y_{\rm PMMA}$ = -2.4 cm. The $x_{\rm PMMA}$ distribution is related to the range of the beam while the $y_{\rm PMMA}$ to its transversal profile. 
Quite remarkably the shape of the distribution of the emission point is approximately the same for protons emitted with different kinetic energies, e.g. the resolution on $x_{\rm PMMA}$ does not depend critically on the $E_{\rm kin}$ variable.%

\noindent
The existence of a relationship between the expected BP position and the peak of the $x_{\rm PMMA}$ distribution, as a function of the PMMA position, in principle could allow us to follow the BP position using the $x_{\rm PMMA}$ measurements.  To estimate the accuracy of this method 
, a position scan has been performed acquiring several data runs moving the PMMA by means of the translation stage.

\noindent
For each run with different PMMA position, the production region of the protons have been monitored using the mean values of the gaussian fit to $x_{\rm PMMA}$ and $y_{\rm PMMA}$ distributions, $\bar{x}_{\rm PMMA}$ and $\bar{y}_{\rm PMMA}$. 
Since $\bar{y}_{\rm PMMA}$ is the coordinate of the proton emission point along the vertical axis, and is related to the fixed beam profile in the transverse plane, its behaviour as a function of the PMMA position provides an estimate of the method's systematic uncertainty.

\begin{figure}[!ht]
\begin{center}
\includegraphics [width = 1 \textwidth] {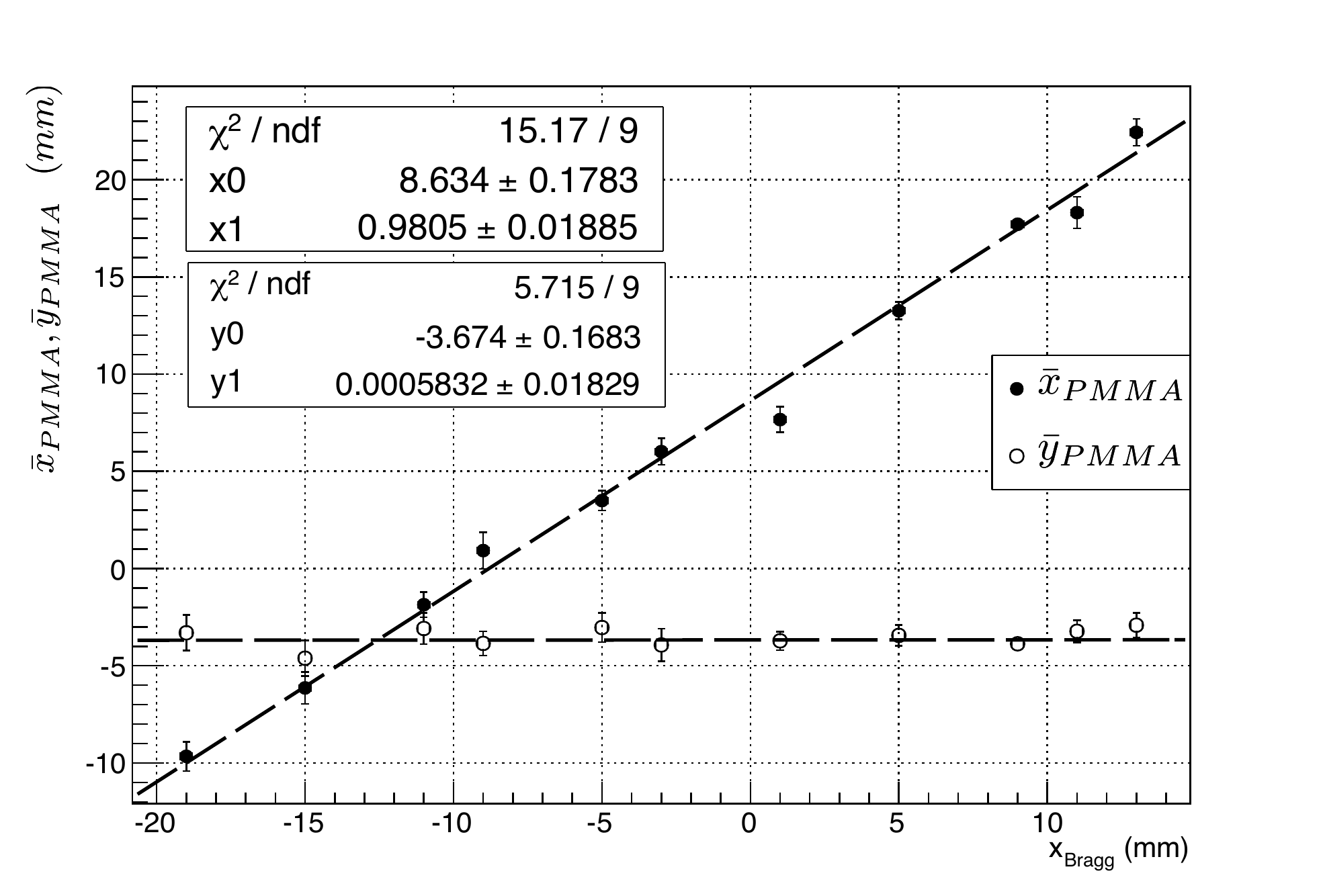}
\caption{\small{Reconstructed peak position of the secondary proton emission distribution $\bar{x}_{\rm PMMA}$,$\bar{y}_{\rm PMMA}$ as a function of the expected Bragg Peak position $x_{\rm Bragg}$, with $E_{\rm kin} >$ 60 MeV.}}
\label{fig:ADD}
\end{center}
\end{figure}

\noindent
Each PMMA position in the stage reference frame can be translated into the expected Bragg peak position $x_{\rm Bragg}$ for that given PMMA position.
Figure~\ref{fig:ADD} shows the results obtained for $\bar{x}_{\rm PMMA}$ and $\bar{y}_{\rm PMMA}$ as a function of $x_{\rm Bragg}$, with $E_{\rm kin} >$ 60 MeV protons. A clear linear relationship  is observed between $\bar{x}_{\rm PMMA}$ and $x_{\rm Bragg}$, indicating that the charged secondary particles emission reconstructed with the drift chamber follows accurately the BP movement.
No dependence of the $\bar{y}_{\rm PMMA}$ values on the Bragg peak position is observed, as expected from a translation of the PMMA along the x-axis only.  
Similar results can be obtained using protons with different $E_{\rm kin}$ selection, as it can be inferred from Figure~\ref{fig:reso_ekin}.

\noindent
To estimate the achievable accuracy on the BP determination several contributions need to be considered.
We evaluated the difference $\Delta_{\rm ProtonBragg} = \bar{x}_{\rm PMMA} - x_{\rm Bragg}$ for all identified protons and for the proton sample with $E_{\rm kin} >$ 60 MeV.  The $\Delta_{\rm ProtonBragg}$ root mean square is $\sigma_{\rm \Delta_{\rm ProtonBragg}}\simeq$ 0.9 mm for both samples.
This can be explained as follows: in the sample with all identified protons the contribution to the total uncertainty due to the scattering is partially compensated by the larger statistics with respect to the sample with $E_{\rm kin} >$ 60 MeV. Table \ref{TAB:NPStat} reports the number of identified protons with $E_{\rm kin} >$ 60 MeV obtained with the position scan data.
\begin{table}[!hbt]
\caption{Statistics of identified protons with $E_{\rm kin} >$ 60 MeV, obtained with the position scan data.}
\begin{center}
\begin{tabular}{cccccccccccc}
\hline
\bs
$x_{\rm Bragg}$ (mm) & -19 & -15 & -11 & -9 &  -5 &  -3 &  1 &    5&    9&    11 &  13 \\
\bs
\hline
\bs
$N^{\rm 60~MeV}_{\rm Protons}$ & 67& 77& 88& 61& 92& 75& 113& 154& 1223& 130& 83 \\
\bs
\hline
\end{tabular}
\label{TAB:NPStat}
\end{center}
\end{table}

\noindent
The uncertainty $\sigma_{\rm Extrapol}$ due to the backward extrapolation of the track from the drift chamber to the beam line can be estimated from the root mean square of the $\bar{y}_{\rm PMMA}$ values, $\sigma_{\bar{y}_{\rm PMMA}} = \sigma_{\rm Extrapol}$ = 0.5 mm. The latter contributes to the $\Delta_{\rm ProtonBragg}$ distribution, together with $\sigma_{\rm Stage}$ = 0.2 mm from the uncertainty on the PMMA positioning. 
We can then estimate the contribution to the total uncertainty coming from the shape of the distribution of the emission point of charged secondary particles as: 
\begin{equation}
\sigma_{\rm Emission} = \sqrt{\sigma_{\rm \Delta_{\rm ProtonBragg}}^2  - \sigma_{\rm Extrapol}^2 - \sigma_{\rm Stage}^2} \sim 0.7 \text{mm}
\end{equation}

\noindent It must be stressed that this value represents only an indication of the precision achievable in the BP determination using secondary protons, due to the target thickness and homogeneity in the present setup, with respect to a possible clinical application.
%
\section{Flux of charged secondary particles}
\label{flux}
The flux of the secondary protons emitted from the beam interaction with the PMMA has been measured at 90$\degree$ with respect to the beam direction and in the geometrical acceptance of the triggering LYSO crystals, configuration maximizing the sensitivity to the Bragg peak position. The surface of the LYSO is 3x3 cm$^2$, corresponding to a solid angle $\Omega_{\rm LYSO} = 1.3\times 10^{-4}$ sr at a distance of 74 cm. The proton's kinetic energy spectrum measured with data has been inserted in the FLUKA simulation to evaluate the detection efficiency in the LYSO crystals for protons with $E_{\rm LYSO} >$ 1 MeV: $\epsilon_{\rm LYSO}= (98.5\pm 1.5)\%$, with the uncertainty mainly due to the Monte Carlo statistics. To properly evaluate the rate of charged secondary particles reaching the LYSO crystals, the number of carbon ions reaching the PMMA target ($N_{\rm C}$) has been computed according to \cite{Agodi2012a}: counting the number of signals in the Start Counter ($N_{\rm SC}$) within randomly-triggered time-windows of $T_w=2\ \micro\second$, corrected for the Start Counter efficiency $\epsilon_{\rm SC} = (96 \pm 1)\%$, and the acquisition dead time. 
The number of emitted secondary protons $N_{\rm P}$ has been measured with the $x_{\rm PMMA}$ distribution counts, corrected for $\epsilon_{\rm SC}$, $\epsilon_{\rm LYSO}$, the tracking efficiency $\epsilon_{\rm Track} = (98 \pm 1)\%$~\cite{Abou-Haidar2012} and the acquisition dead time.

\begin{figure}[!ht]
\begin{center}
\includegraphics [width = 1\textwidth] {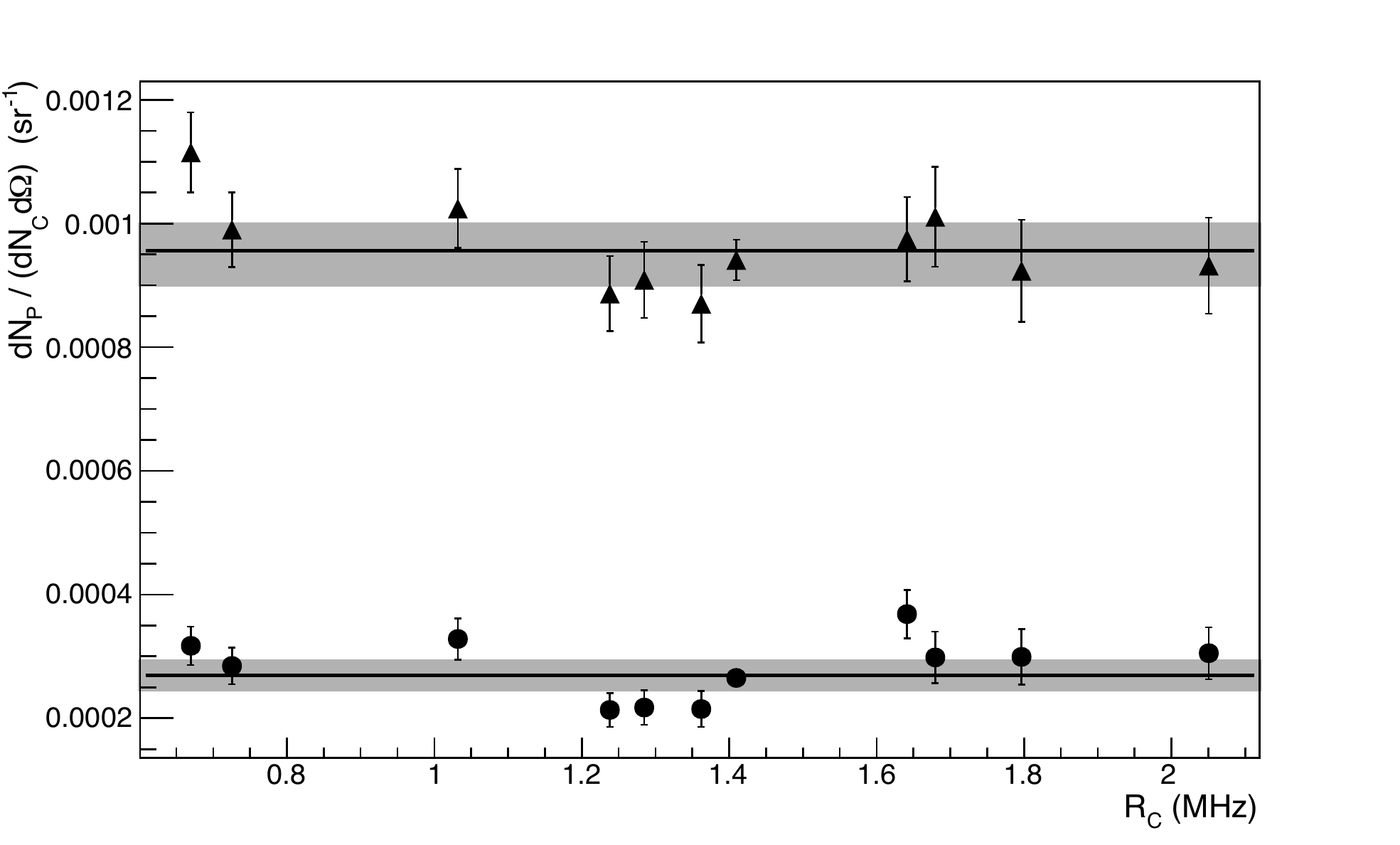}
\caption{\small{Double differential production rate for secondary particles emitted at 90$\degree$ with respect to the beam line, as a function of the rate of the carbon ions $R_{\rm C}$ reaching the PMMA target: all identified protons (triangles) and protons with $E_{\rm kin} >$ 60 MeV (circles). 
}}
\label{fig:flusso}
\end{center}
\end{figure}

\noindent
The double differential production rate of secondary protons emitted at 90$\degree$ with respect to the beam line is estimated as:
\begin{equation}
\frac{d^2N_{\rm P}}{dN_{\rm C}d\Omega}(\theta=90\degree)=\frac{N_{\rm P}}{N_{\rm C}~~\Omega_{\rm LYSO}}.
\end{equation}
Figure~\ref{fig:flusso} shows the double differential production rate of secondary protons, emitted at 90$\degree$ with respect to the beam line, as a function of the rate of the carbon ions $R_{\rm C}$ reaching the PMMA: all identified protons and protons with $E_{kin} >$ 60 MeV.

\noindent Expressing these results in terms of the secondary proton's kinetic energy at emission $E^{\rm Prod}_{\rm kin}$, we obtain:
%
%
%
\begin{eqnarray}
\fl
\frac{dN_{\rm P}}{dN_{\rm C}d\Omega}(E^{\rm Prod}_{\rm kin} > 7 {\rm ~MeV}, \theta=90\degree) =  (9.56\pm 0.18_{\rm stat} \pm 0.40_{\rm sys})\times 10^{-4} sr^{-1} \\
\fl
 \frac{dN_{\rm P}}{dN_{\rm C}d\Omega}(E^{\rm Prod}_{\rm kin} > 83 {\rm ~MeV}, \theta=90\degree) =  (2.69\pm 0.08_{\rm stat} \pm 0.12_{\rm sys})\times 10^{-4} sr^{-1}
\end{eqnarray}
with the systematic contribution mainly due to proton identification and to the uncertainty on the production kinetic energy related to the beam's transversal profile uncertainty. 

\noindent
The same experimental setup described in Section \ref{setup} has been used to measure the differential production rate for prompt photons, with energy $E_{\rm LYSO} >$ 2 MeV and emitted at 90$\degree$ with respect to the beam line: $dN_{\rm \gamma}/(dN_{\rm C}d\Omega)(E_{\rm LYSO} > 2 {\rm ~MeV} , \theta=90\degree) = (2.92 \pm 0.19)\times 10^{-4} sr^{-1}$~\cite{Agodi2012a}.

\section{Discussion and conclusions}
We reported the study of secondary charged particles produced by the interaction of 80 MeV/u fully stripped carbon ion beam of INFN-LNS laboratory in Catania with a PMMA target. 
Protons have been identified exploiting the energy and time of flight measured with a plastic scintillator together with LYSO crystals, and their direction has been reconstructed with a drift chamber. A detailed simulation of the setup based on the FLUKA package has been done to evaluate its acceptance and efficiency, and to optimize secondary particle's identification.
\\
\noindent
It has been shown that the backtracking of secondary protons allows their emission region in the target to be reconstructed. Moreover the existence of a correlation between the reconstructed production region of secondary protons and the Bragg peak position has been observed, performing a position scan of the target.
The achievable accuracy on the Bragg peak determination exploting this procedure has been estimated to be in the submillimeter range, using the described setup and selecting secondary protons with kinetic energy at emission $E^{\rm Prod}_{\rm kin} >$ 83 MeV. 
\\
\noindent
The obtained accuracy on the position of the released dose should be regarded as an indication of the achievable accuracy for possible applications of this technique  to monitor the BP position in hadrontherapy treatment. In fact in clinical application the secondary particles should cross a larger amount of material (patient tissue) resulting in an increased multiple scattering contribution worsening the BP resolution by, at most, a factor 2-3. On the other hand an optimized device allowing a closer positioning to the patient could greatly improve the collected statistics of protons produced with $E^{\rm  Prod}_{\rm kin} >$ 80 MeV, reducing multiple scattering effects. Furthermore the intrinsic good tracking resolution and high detection efficiency easily achievable in charged particles detectors, make this monitoring option worthwhile of further investigations.

\noindent
The measured differential production rate for protons with $E^{\rm Prod}_{\rm kin} >$ 83 MeV and emitted at 90$\degree$ with respect to the beam line is: $dN_{\rm P}/(dN_{\rm C}d\Omega)(E^{\rm Prod}_{\rm kin} > 83 MeV , \theta=90\degree) = (2.69\pm 0.08_{\rm stat} \pm 0.12_{\rm sys})\times 10^{-4} sr^{-1}$.

\ack
We would like to thank the precious cooperation of the  staff of the INFN-LNS (Catania, Italy) accelerator group. The authors would like to thank Dr. M.~Pillon and Dr. M.~Angelone (ENEA-Fra\-scati, Italy) for allowing us to validate the response of our detector to neutrons on the Frascati Neutron Generator; C.~Piscitelli (INFN-Roma, Italy) for the realization of the mechanical support; M.~Anelli (INFN-LNF, Frascati) for the drift chamber construction. 

\noindent
This work has been supported by the ``Museo storico della fisica  e Centro di studi e ricerche Enrico Fermi''. 
\section*{References}
\bibliographystyle{jphysicsBforPMB}
\bibliography{PMBColl}
\end{document}